\documentclass[12pt,hyper]{JHEP3}
\usepackage{graphics,psfrag}
\usepackage{amsmath,amsthm,amssymb,epsfig,euscript,array,cite,cancel}
\usepackage{cases,empheq}
\theoremstyle{definition}

\theoremstyle{remark}

\newcounter{multieqs}

\newcommand{\be}{\begin{equation}}
\newcommand{\ee}{\end{equation}}
\newcommand{\eq}[1]{(\ref{#1})}
\newcommand{\bit}{\begin{itemize}}  \newcommand{\eit}{\end{itemize}}

\newcommand{\bra}[1]{\langle #1|}
\newcommand{\ket}[1]{|#1 \rangle}
\newcommand{\ipr}[2]{\langle #1 | #2 \rangle}

\newcommand{\bm}[1]{\mbox{\boldmath $#1$}}
\newcommand{\rf}[1]{(\ref{#1})}

\def\bd{\begin{document}}
\def\ed{\end{document}}
\def\nn{\nonumber}
\def\bea{\begin{eqnarray}}
\def\eea{\end{eqnarray}}
\let\bm=\bibitem

\def\la{\langle}
\def\ra{\rangle}

\def\npb#1#2#3{Nucl. Phys. {\bf{B#1}} #3 (#2)}
\def\plb#1#2#3{Phys. Lett. {\bf{#1B}} #3 (#2)}
\def\prl#1#2#3{Phys. Rev. Lett. {\bf{#1}} #3 (#2)}
\def\prd#1#2#3{Phys. Rev. {D-\bf{#1}} #3 (#2)}
\def\cmp#1#2#3{Comm. Math. Phys. {\bf{#1}} #3 (#2)}
\def\cqg#1#2#3{Class. Quantum Grav. {\bf{#1}} #3 (#2)}
\def\nppsa#1#2#3{Nucl. Phys. B (Proc. Suppl.) {\bf{#1A}}#3 (#2)}
\def\ap#1#2#3{Ann. of Phys. {\bf{#1}} #3 (#2)}
\def\ijmp#1#2#3{Int. J. Mod. Phys. {\bf{A#1}} #3 (#2)}
\def\rmp#1#2#3{Rev. Mod. Phys. {\bf{#1}} #3 (#2)}
\def\mpla#1#2#3{Mod. Phys. Lett. {\bf A#1} #3 (#2)}
\def\jhep#1#2#3{J. High Energy Phys. {\bf #1} #3 (#2)}
\def\atmp#1#2#3{Adv. Theor. Math. Phys. {\bf #1} #3 (#2)}

\def\N{{\cal N}}
\def\sst{\scriptscriptstyle}
\def\thetabar{\bar\theta}
\def\Tr{{\rm Tr}}
\def\one{\mbox{1 \kern-.59em {\rm l}}}

\def\a{\alpha}      \def\da{{\dot\alpha}}  \def\dA{{\dot A}}
\def\b{\beta}       \def\db{{\dot\beta}}  
\def\g{\gamma}  \def\G{\Gamma}  \def\dc{{\dot\gamma}}  
\def\d{\delta}  \def\D{\Delta}  \def\ddt{\dot\delta}  
\def\e{\epsilon}        \def\ve{\varepsilon}  
\def\f{\phi}    \def\F{\Phi}    \def\vvf{\f}  
\def\h{\eta}  
\def\k{\kappa}  
\def\l{\lambda} \def\L{\Lambda}  
\def\m{\mu} \def\n{\nu}  
\def\o{\omega}  
\def\p{\pi} \def\P{\Pi}  
\def\r{\rho}  
\def\s{\sigma}  \def\S{\Sigma}  
\def\t{\tau}  
\def\th{\theta} \def\Th{\Theta} \def\vth{\vartheta}  
\def\X{\Xeta}  
\def\z{\zeta}  

\def\na{\nabla}  

\def\cA{{\cal A}} \def\cB{{\cal B}} \def\cC{{\cal C}}  
\def\cD{{\cal D}} \def\cE{{\cal E}} \def\cF{{\cal F}}  
\def\cG{{\cal G}} \def\cH{{\cal H}} \def\cI{{\cal I}}  
\def\cJ{{\cal J}} \def\cK{{\cal K}} \def\cL{{\cal L}}  
\def\cM{{\cal M}} \def\cN{{\cal N}} \def\cO{{\cal O}}  
\def\cP{{\cal P}} \def\cQ{{\cal Q}} \def\cR{{\cal R}}  
\def\cS{{\cal S}} \def\cT{{\cal T}} \def\cU{{\cal U}}  
\def\cV{{\cal V}} \def\cW{{\cal W}} \def\cX{{\cal X}}  
\def\cY{{\cal Y}} \def\cZ{{\cal Z}}

\def\ua{\underline{\alpha}}  
\def\uc{\underline{\phantom{\alpha}}\!\!\!\gamma}  
\def\um{\underline{\mu}}  
\def\ud{\underline\delta}  
\def\ue{\underline\epsilon}  
\def\una{\underline a}\def\unA{\underline A}  
\def\unb{\underline b}\def\unB{\underline B}  
\def\unc{\underline c}\def\unC{\underline C}  
\def\und{\underline d}\def\unD{\underline D}  
\def\une{\underline e}\def\unE{\underline E}  
\def\unf{\underline{\phantom{e}}\!\!\!\! f}\def\unF{\underline F}  
\def\unm{\underline m}\def\unM{\underline M}  
\def\unn{\underline n}\def\unN{\underline N}  
\def\unp{\underline{\phantom{a}}\!\!\! p}\def\unP{\underline P}  
\def\unq{\underline{\phantom{a}}\!\!\! q}  
\def\unQ{\underline{\phantom{A}}\!\!\!\! Q}  
\def\unH{\underline{H}}  
  
\def\As {{A \hspace{-6.4pt} \slash}\;}  
\def\bs {{b \hspace{-6.4pt} \slash}\;}  
\def\Ds {{D-\hspace{-6.4pt} \slash}\;}
\def\Gts {{\Gt \hspace{-6.4pt} \slash}\;}
\def\ds {{\del \hspace{-6.4pt} \slash}\;}  
\def\ss {{\s \hspace{-6.4pt} \slash}\;}  
\def\ks {{ k \hspace{-6.4pt} \slash}\;}  
\def\ps {{p \hspace{-6.4pt} \slash}\;}   
\def\xs {{x \hspace{-6.4pt} \slash}\;}  
\def\pas {{{p_1} \hspace{-6.4pt} \slash}\;}  
\def\pbs {{{p_2} \hspace{-6.4pt} \slash}\;}   
\def\cFs {{{\cal F} \hspace{-6.4pt} \slash}\;}

\def\Ah{{\hat{A}}}  
\def\Dh{{\hat{D}}}
\def\Gh{{\hat{G}}}
\def\Fh{{\hat{F}}}
\def\Ih{{\hat{I}}} 
\def\Jh{{\hat{J}}} 
\def\Kh{{\hat{K}}}
\def\Lh{{\hat{L}}} 
\def\Ph{{\hat{P}}}
\def\Rh{{\hat{R}}}
\def\Vh{{\hat{V}}} 
\def\Xh{{\hat{X}}}
 
\def\ah{{\hat{\a}}}
\def\bh{{\hat{\b}}}
\def\gh{{\hat{\g}}}
\def\dh{{\hat{\d}}}
\def\hh{\hat{h}}
\def\uh{\hat{u}}  
\def\xh{\hat{x}}  
\def\yh{\hat{y}}  
\def\ph{\hat{p}}  
\def\xih{\hat{\xi}}  
\def\chih{\hat{\chi}}  
\def\Psih{\hat{\Psi}}

\def\psit{\tilde{\psi}}  
\def\Psit{\tilde{\Psi}}   
\def\Psibt{\tilde{\bar{Psi}}}  

\def\st{\tilde{\sigma}}  

\def\delt{\tilde{\delta}}
\def\Phit{\tilde{\Phi}}   
\def\Phitb{\overline{\tilde{Phi}}}  
\def\tht{\tilde{\th}}  
\def\lt{\tilde{\l}}
\def\chit{\tilde{\chi}}   
\def\phit{\tilde{\phi}} 

\def\At{\tilde{A}}
\def\Bt{\tilde{B}}
\def\Ct{\tilde{C}}
\def\Dt{\tilde{D}}
\def\Et{\tilde{E}}
\def\Ft{\tilde{F}}
\def\Gt{\tilde{G}}
\def\Ht{\tilde{H}}
\def\It{\tilde{I}}
\def\Jt{\tilde{J}}
\def\Qt{\tilde{Q}}  
\def\Rt{\tilde{R}}  
\def\Mt{\tilde{M }}  
\def\Nt{\tilde{N}}   
\def\St{\tilde{S}}
\def\Vt{\tilde{V}}
\def\Xt{\tilde{X}} 
\def\at{\tilde{a}}
\def\ct{\tilde{c}}
\def\dt{\tilde{d}}
\def\htt{\tilde{h}} 
\def\ft{\tilde{f}}
\def\gt{\tilde{g}}
\def\pt{\tilde{p}}  
\def\qt{\tilde{q}}  
\def\vt{\tilde{v}}  
\def\nt{\tilde{n}}  
\def\ut{\tilde{u}}  
\def\wt{\tilde{w}}  
\def\zt{\tilde{z}} 
\def\xt{\tilde{x}} 
\def\yt{\tilde{y}} 
\def\Psit{\tilde{\Psi}}
\def\vphit{\tilde{\varphi}}  

\def\eb{\bar{\epsilon}} 
\def\delb{\bar{\partial}}  
\def\thb{\bar{\theta}}
\def\mub{\bar{\mu}}
\def\lamb{\bar{\l}}
\def\psib{\bar{\psi}}
\def\sb{\bar{\sigma}}
\def\xib{\bar{\xi}}
\def\chib{\bar{\chi}}

\def\Psib{\bar{\Psi}}
\def\Phib{\bar{\Phi}}
\def\Lamb{\bar{\Lambda}}
\def\Sb{{\overline \Sigma}}
\def\cb{\bar{c}}
\def\hb{\bar{h}}
\def\qb{\bar{q}}
\def\wb{\bar{w}}
\def\ub{\bar{u}}
\def\zb{{\bar{z}}}
\def\Hb{\bar{H}}
\def\Qb{{\bar Q}}
\def\Omegab{\overline{\Omega}}
\def\ob{\overline{\omega}}

\def\Ab{{\overline A}} \def\Bb{{\overline B}} \def\Cb{{\overline C}}  
\def\Db{{\overline D}} \def\Eb{{\overline E}} \def\Fb{{\overline F}}  
\def\Gb{{\overline G}} 
\def\Ib{{\overline I}}  
\def\Jb{{\overline J}} \def\Kb{{\overline K}} \def\Lb{{\overline L}}  
\def\Mb{{\overline M}} \def\Nb{{\overline N}} \def\Ob{{\overline O}}  
\def\Pb{{\overline P}}  \def\Rb{{\overline R}}  
 \def\Tb{{\overline T}} \def\Ub{{\overline U}}  
\def\Vb{{\overline V}} \def\Wb{{\overline W}} \def\Xb{{\overline X}}  
\def\Yb{{\overline Y}} \def\Zb{{\overline Z}}  

\def\fb{{\overline f}}
\def\gb{{\overline g}}
\def\mb{{\overline m}}
\def\lb{{\overline l}}
\def\yb{{\overline y}}
  
\def\ldel{{\overleftarrow{\del}}}
\def\rdel{{\overrightarrow{\del}}}
\def\ldeldel{{\overleftarrow{\del^2}}}
\def\rdeldel{{\overrightarrow{\del^2}}}
\def\ldelb{{\overleftarrow{\bar{\del}}}}
\def\rdelb{{\overrightarrow{\bar{\del}}}}

\def\ba{{\bf a}} 
\def\bk{{\bf k}}  
\def\bl{{\bf l}}  
\def\bp{{\bf p}}  
\def\bq{{\bf q}}  
\def\br{{\bf r}}
\def\bt{{\bf t}}
\def\bu{{\bf u}}
\def\bv{{\bf v}}
\def\bx{{\bf x}}  
\def\by{{\bf y}}  
\def\bR{{\bf R}}  
\def\bV{{\bf V}}

\def\bone{{\bf 1}}  

\def\va{{\vec a}}
\def\vk{{\vec k}}
\def\vp{{\vec p}}
\def\vq{{\vec q}}
\def\vx{{\vec x}}
\def\vy{{\vec y}}
\def\vu{{\vec u}}
\def\vv{{\vec v}}

\def\vs{{\vec \sigma}}
\def\vtau{{\vec \tau}}

\newcommand{\ov}[1]{\overrightarrow{#1}}

\def\frA{\mathfrak{A}}
\def\frB{\mathfrak{B}}
\def\frC{\mathfrak{C}}
\def\frD{\mathfrak{D}}
\def\frE{\mathfrak{E}}
\def\frF{\mathfrak{F}}
\def\frG{\mathfrak{G}}
\def\frH{\mathfrak{H}}
\def\frM{\mathfrak{M}}
\def\frN{\mathfrak{N}}
\def\frR{\mathfrak{R}}
\def\frW{\mathfrak{W}}

\def\fra{\mathfrak{a}}
\def\frb{\mathfrak{b}}
\def\frf{\mathfrak{f}}
\def\frg{\mathfrak{g}}
\def\frh{\mathfrak{h}}
\def\frl{\mathfrak{l}}
\def\frs{\mathfrak{s}}
\def\fri{\mathfrak{i}}
\def\frj{\mathfrak{j}}

\def\ma{\mathfrak{a}}
\def\mg{\mathfrak{g}}
\def\mh{\mathfrak{h}}
\def\mR{\mathfrak{R}}
\def\mN{\mathfrak{N}}

\def\d{\delta}\def\D{\Delta}\def\ddt{\dot\delta}  
  
\def\pa{\partial} \def\del{\partial}  
\def\xx{\times}  
\def\uno{\mbox{1 \kern-.59em {\rm l}}}    
  
\def\trp{^{\top}}  
\def\inv{^{-1}}  
\def\dag{{^{\dagger}}}  
\def\pr{^{\prime}}  
  
\def\rar{\rightarrow}  
\def\lar{\leftarrow}  
\def\lrar{\leftrightarrow}  
  
\newcommand{\0}{\,\!}      
\def\one{1\!\!1\,\,}  
\def\im{\imath}  
\def\jm{\jmath}  
  
\newcommand{\tr}{\mbox{tr}}  
\newcommand{\slsh}[1]{/ \!\!\!\! #1}  
  
\def\vac{|0\rangle}  
\def\lvac{\langle 0|}  
  
\def\hlf{\frac{1}{2}}  
\def\ove#1{\frac{1}{#1}}  

\def\Box{\square}  
\def\CC {\mathbb{C}}
\def\FF {\mathbb{F}}
\def\RR{\mathbb{R}}
\def\NN{\mathbb{N}}  
\def\ZZ{\mathbb{Z}}  
\def\bb#1{{\bf #1}}  
\def\bcomment#1{}  
\def\bfhat#1{{\bf \hat{#1}}}  
\def\VEV#1{\left\langle #1\right\rangle}  

\newcommand{\ex}[1]{{\rm e}^{#1}} \def\ii{{\rm i}}  

\newcommand{\lrbrk}[1]{\left(#1\right)}
\newcommand{\sfrac}[2]{{\textstyle\frac{#1}{#2}}}
 
\def\stw{{\sqrt{2}}}

\def\rf {{\rm f}}
\def\ri {{\rm i}}
\def\rj {{\rm j}}
\def\rk {{\rm k}}
\def\rl {{\rm l}}
\def\rs {{\scriptscriptstyle \rm S}}
\def\rt {{\scriptscriptstyle \rm T}}

\def\rQ {{\scriptscriptstyle \rm \cQ}}
\def\rR {{\scriptscriptstyle \rm \cR}}

\def\cQb{{\cal \Qb}}
\def\cRb{{\cal \Rb}}
\def\cWb{{\cal \Wb}}

\def\fd {{\rm N}}
\def\afd {{\overline{\rm N}}}

\def \II {I\hspace{-.1em}I\hspace{.1em}}
\def \IIA {\mbox{\II A\hspace{.2em}}}
\def \IIB {\mbox{\II B\hspace{.2em}}}
\def \gs {g^s}
\def \ls {\lambda^s}

\def \I {{\cal I}}
\def \qs {q\hspace{-.53em}/\hspace{.15em}}
\def \ks {k\hspace{-.53em}/\hspace{.15em}}
\def \YM {{\mbox{\tiny YM}}}
\def \gym {g_{\YM}}

\def \Lc {\L_c}
\def\IR{\relax{\rm I\kern-.18em R}}
\def \id {{\bf 1}}

\def\cci{\ell}
\def\ccj{\ell'}

\def \thbb{\overline{\th\th}}
\newcommand \ol{\overline}
\def \lamb{\bar{\lambda}}
\def \vphi{\varphi}
\def \lambh{\hat{\bar{\lambda}}}
\def \lh{\hat{\lambda}}
\def \dd{\ddagger}

\def \Xd{\dot{X}}

\author{Chong-Sun Chu and Gurdeep S. Sehmbi\\  
Centre for Particle Theory
and Department of Mathematical Sciences,\\ 
Durham University, Durham, DH1 3LE, UK \\
E-mail:  
\email{chong-sun.chu@durham.ac.uk}, \email{g.s.sehmbi@durham.ac.uk} }

\title {
D1-Strings in Large RR 3-Form Flux, \\
Quantum Nambu Geometry and \\
M5-Branes in $C$-Field
}

\abstract{
We consider D1-branes in a RR flux background and show that there is a
low energy - large flux double scaling limit 
where the D1-branes action is dominated by a
Chern-Simons-Myers coupling term.
As a classical solution to the matrix model,
we find a novel quantized geometry 
characterized by a quantum Nambu
3-bracket. 
Infinite dimensional representations of the 
quantum Nambu geometry are constructed which demonstrate that
the quantum Nambu geometry is
intrinsically different from the ordinary Lie algebra type 
noncommutative geometry. 

Matrix models 
for the \IIB  string, \IIA string and M-theory in 
the corresponding backgrounds  are constructed. 
A classical solution of
a quantum Nambu
geometry in the \IIA Matrix string theory gives rise to an expansion of 
the fundamental strings into a system of multiple D4-branes and the
fluctuation 
is found to describe an action for a non-abelian
3-form field strength which is a natural  non-abelian generalization of 
the  PST action for a single D4-brane. 

In view of the recent proposals
\cite{d1,d2} of the M5-branes theory in terms of the D4-branes, 
we suggest a natural way to
include all the KK modes and propose 
an action for the the multiple M5-branes in a constant $C$-field.
The worldvolume of the M5-branes in a $C$-field is found to be described by a
quantum Nambu geometry with self-dual parameters.
It is intriguing that our action is naturally 
formulated in terms of a 1-form gauge field 
living on a six 
dimensional 
quantum Nambu geometry.

}
\preprint{DCPT-11/45}
\keywords{M-Theory, D-branes, M-branes, Nambu bracket}

\begin{document}

\section{Introduction}

It is generally expected that the usual description of
spacetime in terms of Riemannian geometry would break down above the
Planck energy scale. A possibility is that geometry is quantized and 
spacetime coordinates become quantum operators. In this case, traditional 
spacetime concepts such as locality and causality, 
and
even the fundamental nature of spacetime itself, 
will have to be re-examined. 
String theory, 
as a candidate for a theory of quantum gravity, provides an interesting 
setup to address some of these questions.  
One of our  motivations is to discover  
new types of quantum geometries in string theory and 
to study the physics on these quantum
spaces. 

There are a number of ways where
a quantum geometry could emerges in string theory. 
One way is to consider
open string theory ending on a D-brane with a background $B$-field on
it  and use the open string to probe the 
geometry of the D-brane. The obtained form of the noncommutative geometry
could be of Moyal type \cite{dh,CH,sch1,sw}, 
\be \label{ncg}
[X^\m,X^\n] = i \th^{\m\n},
\ee
or a fuzzy sphere \cite{sch2,sch3}. String effects to arbitrary loops can
be included readily \cite{loop}. 
Noncommutative geometry could also
arise in matrix models \cite{bfss,ikkt} as a classical solution 
\cite{bss,bkgd-indep}, \cite{makeenko,smith-2b}.
Myers effects \cite{myers} could introduce additional terms to the
matrix model and lead to new solutions 
\cite{susskind,iso-fuzzy,BMN}.
We note that 
all these quantized geometries are characterized by a
commutator and could be referred to as of Lie-algebra type.
Inclusion of small fluctuations
around these solutions always gives rise to a noncommutative gauge theory 
\cite{bkgd-indep}. The discovery of
a new non Lie-algebra type of quantum geometry in string theory would be
interesting
\footnote{We remark that very interestingly, nonassociative geometry
has also made an appearance in string theory recently \cite{nonass}. 
Unlike the example \cite{sch2}
where the nonassociativity is due to a projection of the spectrum, here 
it seems the emergence of the nonassociativity is due to the
insistence of the use
of a geometric language in a non-geometric background. In this paper
we are interested  in quantum geometries that are characterized 
by conventional associative operators. 
We thank Erik Plauschinn for a discussion on this.
}
. Moreover, one may wonder if the physics of the fluctuations may 
lead to some new kind of gauge theory. This is another
motivation for this work.

This  aim has been achieved partially in \cite{CS1}
where it was shown
that the
consistency between the different descriptions of the M2-M5 
intersecting branes
system implies that the M5-brane geometry 
in the presence of a constant 3-form $C$-field takes the form of 
\be \label{3bkt-geom}
[X^\m,X^\n,X^\l] =i\th^{\m\n\l},
\ee
where $\th$ is a constant and the 3-bracket is given by a Lie 3-bracket
\footnote{
In \cite{CS1}, it was also
shown that the standard
noncommutative geometry \eq{ncg} of D-branes in 
a constant 2-form $B$-field could be derived similarly by
considering the intersecting system of F1 strings and D3 brane in
$B$-field. In this case, the 2-bracket $[ \cdot,\cdot]$ is a Lie
2-bracket and is inherited from the Lie 2-bracket structure 
of the boundary matrix string theory.}.
A Lie 3-bracket 
is multilinear and is antisymmetric under 
interchange of any pair of its components. Moreover it satisfies 
the fundamental identity
\be \label{FI}
[[f,g,h],k,l] = [[f,k,l],g,h]+ [f,[g,k,l],h]+ [f,g,[h,k,l]], 
\ee
where $f,g,h,k,l$ are any  elements of the algebra.  
The reason why a Lie 3-bracket appears is because 
the geometry of the M5-brane was 
inferred from the boundary dynamics 
of the open M2-branes which end on it; and the BLG model
\cite{BLG} with boundary 
was used to 
describe the open M2-branes.
In a quantum theory, it is
necessary to understand the relation \eq{3bkt-geom} as an operator 
relation. However, the representation of the
Lie 3-algebra relation as transformations on vector spaces 
or  
maybe 
some kind of generalization is still an open question
(see \cite{sz,cw} for some different approaches). 
Since the
difficulties are mainly due to 
the insistence of the fundamental identity, it motivates us
to look for a 3-bracket geometry of the form \eq{3bkt-geom} 
but
with a 3-bracket
where the fundamental identity is not required
\footnote{One may think that one could easily repeat 
the analysis of \cite{CS1} with the open ABJM theory \cite{ABJM}
instead and obtain a 
similar relation \eq{3bkt-geom} where $X^i$'s would be operators. 
This is however
not immediate
since the analysis in \cite{CS1} involves a comparison of
the results obtained in different dual descriptions of the M2-M5
intersecting branes system. In particular a comparison of the
information
contained in the boundary condition of the open M2-branes system and
those in the  3-sphere (or 3-ellipsoid in the presence of $C$-field) 
description of the
M2-branes spike is needed. However this is tricky for the ABJM
theory since classically only a fuzzy two sphere is seen, see 
\cite{ram} for a careful analysis on this issue.
}
. 

We remark  that although the fundamental identity 
plays a crucial role in the  BLG model,
since it allows gauge transformations to be defined
in terms of the 3-bracket and
ensures the closure of the supersymmetry algebra, a priori 
there is no reason that a quantum geometry of the 3-bracket form
\eq{3bkt-geom} 
should obey the fundamental identity. In
particular if a 3-bracket geometry of the form \eq{3bkt-geom} 
has a different physical origin which is not related to M2-branes, 
then one would not expect the fundamental identity to
be observed.
In this paper, we will show that the quantum 
geometry \eq{3bkt-geom}
arises as a classical solution of a matrix model of D1-strings in
a background of large RR 3-form flux. Here the 3-bracket 
is given by
one  which is defined on ordinary operators:
\be \label{n-3bkt}
[f,g,h] := fgh + ghf +hfg - fhg-gfh-hgf,
\ee
where $f,g,h$ are any three operators and the binary product is the usual 
operator product. It is easy to see that the fundamental identity is
not observed by 
the 3-bracket \eq{n-3bkt}, see \cite{zachos} for a discussion of this
as well as some other algebraic properties of the 3-bracket \eq{n-3bkt}.  
In fact if one were to
give up the fundamental identity, the most natural 3-bracket to
consider would be the one given by \eq{n-3bkt} which is the most natural higher
order  generalization of the  commutator. 
The 3-bracket \eq{n-3bkt} was
originally introduced by Nambu \cite{nambu} as a possible
candidate of the quantization of the classical Nambu bracket 
$ \{f,g,h\}
: = \e^{ijk} \del_i f \del_j g \del_k h$. 
Therefore we will refer to \eq{n-3bkt} as the {\it quantum Nambu  bracket} 
and the
geometry \eq{3bkt-geom} as the {\it quantum Nambu geometry}.
 
For the
standard 
noncommutative geometry \eq{ncg}, a 2-form field
strength can be written as 
\be
F^{\m\n} = -i [X^\m,X^\n]
\ee 
when the fluctuation over the noncommutative geometry background is taken 
into account. 
What about the fluctuation around the quantum Nambu geometry?
It is suggestive to interpret the quantum
Nambu bracket of the target space coordinate fields $X^\m$ 
as a 3-form field strength
\be \label{H-id}
H^{\m\n\l} = -i [X^\m,X^\n, X^\l].
\ee
To check this idea, we have to look for a place where a non-abelian
3-form field strength lives. This leads us to consider the system of 
multiple D4-branes (where the 3-form field strength would be the
Hodge dual of a 2-form field strength) 
and multiple M5-branes (where the 3-form
field strength would be self-dual). 

To reach the D4-branes system, 
we use the D1-strings matrix model to derive the
Matrix model descriptions for the \IIB string, \IIA string and M-theory in 
a corresponding background of large RR-flux or its uplift to eleven dimensions. 
For the \IIA Matrix string theory, we find that a classical solution
of quantum Nambu geometry is again allowed. We also find that 
the fluctuation around the solution gives a Lagrangian for a 1-form
gauge potential whose form is 
exactly the same as 
the  dimensional reduced PST action (which describes a single D4-brane) 
\cite{pst} if
a quantum Nambu bracket of $X^\m$ 
is identified as \eq{H-id} as a 3-form field strength whose Hodge dual would
be the Yang-Mills field strength. Physically, this means 
the system of fundamental strings has expanded over the quantum Nambu
geometry into a system of multiple D4-branes. 

Since a system of multiple D4-branes can be
considered as a dimensional reduction of multiple M5-branes on a circle,
it has been proposed recently that \cite{d1,d2} the instantons on
the D4-branes can be identified with the KK modes of the compactified
M5-branes, and that by including all the instantons, the D4-branes
SYM theory is in fact equal to the M5-branes theory. 
In view of this, we suggest a natural way to
include all the KK modes into the D4-branes and propose the
action \eq{SM55} for multiple M5-branes in a constant self-dual $C$-field. 
Our proposed action is living on a 6-dimensional 
quantum Nambu geometry with self-dual parameter $\th^{\m\n\l}$
and is formulated in terms of a non-abelian 
3-form field strength 
defined using \eq{H-id}. A priori, such an $H^{\m\n\l}$ may not obey
the desired self-duality condition. 
Nevertheless quite amazingly 
we find that the self-duality condition emerges naturally from our model.
The M5-branes system in a $C$-field could be reduced to a system of 
D4-branes in $B$-field, and the latter has a worldvolume described by
the standard Moyal type noncommutative geometry. This connection allows
us to identify the $\th^{\m\n\l}$  parameter of the
quantum Nambu geometry  as a $C$-field on the worldvolume of the
M5-branes. Therefore we obtain the result that the 
worldvolume of the M5-branes in a $C$-field is 
described by a
quantum Nambu geometry 
\be
[X^\m,X^\n,X^\l] = i \th^{\m\n\l},
\ee
with self-dual parameters $\th^{\m\n\l} = C^{\m\n\l}$.

The plan  of the paper is as follows. 
In section 2, we consider D1-branes in a 
constant RR 3-form flux background. We show that there is a low energy
- large
flux double scaling limit such that the D1-branes action is dominated
by the RR coupling terms. 
We then show that the resulting D1-branes matrix model has the
quantum Nambu geometry as a 
classical solution. 
In
section 3, we present some  analysis of the mathematical properties 
of the quantum Nambu geometry. Infinite dimensional representations
are constructed, and we explain how the existence of these
representations implies that the quantum Nambu geometry is
intrinsically  different
from the ordinary Lie algebra type geometry. 
In section 4, we derive the
Matrix model descriptions for the \IIB string, \IIA string and M-theory in
the corresponding backgrounds. 
In section 5, we argue and propose the action \eq{SM55} as the action
for a system of multiple M5-branes in a constant $C$-field. The
worldvolume geometry of the system of M5-branes is argued to be given
by a quantum Nambu geometry with self-dual parameters   $\th^{\m\n\l}
= C^{\m\n\l}$.
In our formulation, the fundamental dynamical variables is a 1-form gauge 
potential and the 3-form field strength is constructed out of them as a
Nambu bracket. We discuss and give comments on this dual formulation.
The paper is concluded with some further discussions.

\section{Matrix Model of D1-Strings in Large RR 3-Form Flux}

\subsection{A \IIB supergravity background}
 
In the paper \cite{CH-nambu}, an exact \IIB supergravity background with a 
constant  RR 3-form flux  was constructed. 
The background was constructed by turning on a constant RR 3-form flux  
in the $AdS_5$ factor of the  standard  $AdS_5 \times S^5$ background.
The background has a spacetime which is a direct product
\be
\cM = \cM_5 \times \cM_5'
\ee
and has a nonvanishing dilaton, axion,
RR potentials $C_2$ and $C_4$ specified by:
\bea
e^{-\Phi} &=& \chi /(2 \sqrt{2}) = \mbox{constant}, \\
F_3 &=&
\begin{cases}
f \e_{ijk}, & i,j,k =1,2,3, \\
0, & {\rm otherwise},
\end{cases} \label{F-ans}\\
F_5 &=& \begin{cases} 
c \ve_5 & \mbox{on $\cM_5$}, \\
c \ve'_5 & \mbox{on $\cM_5'$}, \\
0 & \mbox{otherwise}.
\end{cases} 
\eea
In the above, $f$ and $c$ are constants,  
$\ve_5$ and $\ve_5'$ are the volume forms on $\cM_5 (\mu =0,1,2,3,4)$ 
and $\cM_5' (\mu = 5,6,7,8,9)$ 
\be
\ve_{\m_0 \cdots \m_4} = \sqrt{-\det G_5} \; \e_{\m_0 \cdots \m_4}, \qquad
\ve'_{\m_5 \cdots \m_9} = \sqrt{-\det G'_5} \; \e_{\m_5 \cdots \m_9}, 
\ee 
and  $\e_{\m_0 \cdots \m_4}$, $\e_{\m_5 \cdots \m_9}$  
are the Levi-Civita symbols: 
$\e^{01234} = - \e_{01234} =1, \e^{56789} =\e_{56789} =1$. 

It was found that a consistent background can be constructed if the 
magnitudes of the RR potentials $C_2$ and $C_4$ are chosen appropriately,
\be
f^2 = \frac{2}{3} c^2.
\ee
The metric (in the string frame) takes the form of
$\RR^3 \times AdS_2 \times S^5$:
\be \label{metric}
ds^2 = \sum_{i=1}^3 (dX^i)^2 + R^2 (\frac{-dt^2 +dU^2}{U^2})
+ R'{}^2d \Omega_5^2,
\ee
where 
\be
R^2 = 2e^{-2\Phi}/f^2, \qquad  
R'{}^2 = 80 e^{-2\Phi}/f^2,
\ee
and 
$d \Omega_5^2 = \hat{G}_{i'j'} dX^{i'} d X^{j'}$ 
is the metric for an $S^5$ of unit radius.
Apart from the $\RR^3$ part, the  spacetime can be understood 
as a warping of a one-dimensional Minkowski space $M_1$ with  a six-dimensional
manifold $Y_6$ with a conical singularity at $U=0$.

For later use, we record  the 
RR 2-form potential 
\be \label{C-ans}
C_2 = f \e_{ijk} X^i dX^j dX^k, 
\qquad i,j,k =1,2,3.
\ee
Our convention is 
$F_{\m\n\l} = \frac{1}{3}(\del_\m C_{\n\l} + \del_\n C_{\l\m} + \del_\l C_{\m\n})$.
The expressions for $C_4$ is more complicated. 
Later we will consider a large $f$ limit for a system of 
D1-strings in this background. 
For our purpose, it is enough to 
note that
$C_{\m_1 \cdots \m_5}, (\mu_i = 0,1, \cdots, 4)$ is proportional to $1/f$ and 
$C_{\m_1 \cdots \m_5}, (\mu_i = 5, 6, \cdots, 9)$ 
is proportional to  $c R^5 \sim 1/f^4$.
We remark that the background is nonsupersymmetric.

\subsection{Matrix model of D1-strings in limit of large $F_3$}

Let us consider a system of $N$ parallel D1-branes in this background.
The worldvolume action for the D1-branes is given by the Non-abelian
Born-Infeld action plus the Chern-Simons term of the Myers type given by 
\cite{myers}
\be
S_{CS} = \m_1\int \Tr P(e^{i \l \ri_\Phi \ri_\Phi} 
\sum_n C_n )e^{\l F}.
\ee
Here $\mu_1 = 1/(g_s 2 \pi \a' )$, $\l= 2 \pi \a'$ and $X^I=2 \pi \a' \Phi^I$.
Our background has $B=0$. With 
$C_2$ and $C_4$ turned on, the Chern-Simons term reads
\bea \label{S-CS}
S_{CS} &=& \m_1 \int  \Tr \left[  \l F \chi +
P\,C_2 + i \l^2 F \ri_\Phi \ri_\Phi C_2 + 
i \l P\,  \ri_\Phi \ri_\Phi C_4 - \frac{\l^3}{2} F \ri_\Phi^4 C_4 
\right] \\
&:=& S_{\chi} + S_{C_2} + S_{C_4}, \label{S-Cs}
\eea
where $S_\chi, S_{C_2}, S_{C_4}$ denote 
the terms in $S_{CS}$ that depend on the RR-potentials $\chi, C_2$ and $C_4$
respectively.
Substituting \eq{C-ans}, we obtain
\bea \label{S-C2}
S_{C_2} /\m_1 &= & f \int  d^2 \s \Tr (
\frac{1}{2} \e_{ijk} X^i D_\a X^j D_\b X^k \e^{\a\b}) 
+ 
 f\int  d^2 \s \Tr (
i  F X^i X^j X^k \e_{ijk})\nn\\
&\equiv&  
f\int d^2\s\, (L_1 + L_2),
\eea
here $\e^{01} = -\e_{01}=1$, 
$F=F_{01}$. From now on we will use $F$ to refer to either 
the curvature two-form or the component $F_{01}$.
It should be clear from the context which is which. 
Naively, if we take a large $F_3$ limit, then  the D1-branes action 
is dominated by $S_{CS}$. This is what we would like to 
demonstrate now. More precisely, we will show that there is a certain double
scaling limit wherein the dynamics of the system of D1-branes is 
dominated by the 
$C_2$ coupling term $S_{C_2}$. 
To do this, we need to include the Non-abelian Born-Infeld action,
examine the large $f$ limit of the 
equations of motion and keep the parts of the action that contribute
in the limit. 

The Non-abelian Born-Infeld theory in curved space is however not so well 
understood. 
First, for flat space, one may expand the Non-abelian 
Born-Infeld action in the powers of $F$. 
However there is an ambiguity associated with 
the ordering of $F$ which cannot be fixed with a simple 
symmetrization procedure  \cite{NBI}, \cite{HT}. 
This ambiguity associated with the ordering of $F$ disappears in the Yang-Mills 
limit.
However in a curved space, 
there is new difficulty associated with the incorporation 
of a curved metric. A natural proposal \cite{curved-d} is 
to promote the metric to become a matrix
$G_{IJ}(X)$ and to incorporate the effect of curved space with the action,
for the case of D1-branes reads,
\bea \label{S-X}
S_{X}/\m_1 := \int d^2 \s \sqrt{-\det G_{\a\b}} \Big(&&
 G_{IJ}(X) D_\a X^I D_\b X^J G^{\a\b} \nn\\
&&+ \frac{1}{\a'} G_{IJ}(X) G_{KL}(X)[X^I,X^K] [X^J,X^L]
\Big),
\eea
where a physical gauge $X^\a =\s^\a$ 
has been taken by making use of the worldvolume 
diffeomorphism of the Non-abelian Born-Infeld theory.
The action \eq{S-X} 
is highly ambiguous due to the ambiguity in the ordering of $X$ in the metric
matrix function $G_{IJ}(X)$. 
For the case of $p=0$, it was proposed that \cite{curved-d} 
the action \eq{S-X} gives the Matrix
theory in curved space and it was found that a large class (but not all) of 
the ambiguities
could be resolved by requiring that the
IR gravitational physics to be correctly reproduced. A more general principle 
is still needed to construct the action unambiguously in general.

Fortunately we will see that these ambiguities will not bother us. 
Let us assume that in the small $\a'$ limit,
the system of D1-branes is described by
an action of the form \eq{S-X} with $I,J = 2,3,\cdots, 9$ together with 
the Chern-Simon coupling \eq{S-CS}. Note that for our metric, 
the ambiguity of the action $S_X$ 
is concentrated entirely in the $S^5$ part. 
The full D1-strings action is given by
\be
S_{D1}:= S_X+ S_{CS} + S_{YM},
\ee 
where the Yang-Mills term is
\be
S_{YM}/\m_1 =  \a'{}^2 
\int \sqrt{-\det{G_{\a\b}}} \; F_{\a\b}F_{\a'\b'} G^{\a\a'} G^{\b\b'}
\ee
and the metric is 
\bea
G_{\a\b} &=& R^2 \s^{-2} \eta_{\a\b}, \qquad\quad\qquad \;\; \a,\b =0,1,\\
G_{ij} &=& \d_{ij} , \qquad\qquad\qquad\qquad\quad i, j =2,3,4, \\
G_{i'j'} &=& R'{}^2 \times \hat{G}_{i'j'}(X^{k'}),
\qquad\quad\qquad i',j' =5,6,7,8,9,
\label{G3}
\eea
with $\hat{G}_{i'j'}$ being the metric for a unit 5-sphere.

It is not difficult to see  that:
\begin{enumerate}
\item the scalars $X^i$ and $X^{i'}$ decouple from each other 
in the action 
$S_{D_1}$.
\item  
the contributions to the equations of motion 
of $X^i$ and $X^{i'}$ from the various pieces of the actions 
\eq{S-CS} and \eq{S-X} are given by: 
\be
\begin{tabular}{|c|cccc|}
\hline
\mbox{contribution of:} & $S_X$  & $S_{C_2} $ 
& $\ri_\Phi^2 C_4$ &  $\ri_\Phi^4 C_4$ \\
\hline
\mbox{EOM of $X^i$:} & $ O(1/\a') $ &  $O(\frac{f}{\a'}) $ 
&  $O(\frac{1}{f\a'{}^2}) $ &  $0 $ \\
\mbox{EOM of $X^{i'}$:} & $ O(\frac{1}{f^2 \a'}) $ &  $0 $ 
&  $O(\frac{1}{f^4 \a'{}^2})  $ &  $O(\frac{1}{f^4 \a'{}^2}) $\\
\hline 
\end{tabular}.
\ee 
\item the equation of motion of $X^{i'}$ can be solved
with $X^{i'} =0$.
\end{enumerate}
These are truly independent of the 
ambiguity of the form of the metric $G_{i'j'}$ 
in  the action \eq{S-X}. 

Therefore, one can set  $X^{i'} =0$ and focus on the sector 
with only the scalars $X^i$ and the gauge field activated. Now the action
$S_{C_2}$ is of order $O(f/\a')$ and the piece of action $\ri_\Phi^2 C_4$ 
in \eq{S-CS} 
is of order $O(1/f\a'^2)$. 
Therefore 
if we take a double scaling limit $\e\to 0$:
\bea \label{scaling}
\a' &\sim& \e, \nn \\
f &\sim& \e^{-a}, \quad a>0,
\eea
such that $a>1/2$, then $S_{C_2}$ dominates. Moreover, 
$S_{YM}$ can be ignored compared to $S_{C_2}$ if $a<2$. All in all, in the 
double scaling limit \eq{scaling} with $1/2 <a<2$, the low energy action of
$N$ D1-branes in a large $F_3$ background is given by
\be
\lim_{\e \rightarrow 0} S_{D1} = S_{C_2}.
\ee
 
We remark that the dominance of the system by a topological term is 
similar to what happened in the discussions of \cite{susskind}, 
where  the effects of a  Lorentz force term 
\be
L = \frac{\m_0 H_3}{2} \ve_{ij}\Tr X^i D_t X^j, \quad i=1,2,
\ee 
on the physics of a system of $N$ D0-branes dissolved in 
a D2-brane (whose spatial
directions are $i=1,2$)
was studied. 
There it was found that 
the equation of motion of $L$ is solved with any time independent
configuration $D_t X^i =0$
and a  
specific solution $[x^i,x^j] = i \th \e^{ij}$ which corresponds to a D2-brane
non-vanishing charge density were considered.

\subsection{Quantum Nambu geometry as classical solution}

We can now find the equations of motion to the action $S_{C_2}$, these are
\bea
&&\ve^{\a\b}\ve_{ijk}[X^j,D_\b X^kX^i] + \ve^{\a\b}\ve_{ijk}
[D_\b, X^iX^jX^k] =0, \\
&&\frac{3}{2}\ve_{ijk}D_\a X^j D_\b X^k \ve^{\a\b} +\ve_{ijk}[F;X^j,X^k]' =0,
\eea
where $[A;B,C]' := [B,C]A + A[B,C] +BAC-CAB$ is antisymmetric only 
in exchange of $B, C$. This bracket arises since 
$\Tr [A,B,C]D = \Tr [D;B,C]' A$, in analogy to the relation
$\Tr D[A,B] = \Tr [D,A] B$ which is useful in ordinary Yang-Mills theory.

The first equation is solved with 
any (covariantly) constant configuration
\be \label{soln1}
D_\a X^i=0.
\ee
The second equation becomes
$ \ve_{ijk}[F,X^j,X^k]'=0$
and is solved by 
\be \label{soln2}
F=0.
\ee
Certainly the standard noncommutative geometry
\be
[X^i,X^j] = i \th^{ij}
\ee
is allowed, but there is also a new solution
\be \label{xxx}
[X^i,X^j,X^k] = i \th \e^{ijk},
\ee
where $\th$ is a constant and the 3-bracket is given by \eq{n-3bkt}. 
We note that
the solution \eq{xxx} is not allowed in the standard matrix models 
\cite{bfss,ikkt}
where  no external $F_3$ is turned on.

We remark that the 3-bracket \eq{n-3bkt} was
originally introduced by Nambu \cite{nambu} as a possible
candidate of the quantization of the classical Nambu bracket 
\be 
\{f,g,h\}
: = \e^{ijk} \del_i f \del_j g \del_k h.
\ee 
In his paper, Nambu was interested in generalizing the Hamiltonian
mechanics to the form (Nambu mechanics)
\be \label{N-mech}
\frac{d f}{d t} = \{ H_1, H_2, f \},
\ee
which involves two ``Hamiltonians'' $H_1, H_2$. The concept of a 
fundamental identity
was not considered in his consideration.
In fact one can easily check that 
the fundamental identity is not satisfied for \eq{n-3bkt}.
The concept of 
fundamental identity was introduced almost 20 years later by
Takhtajan \cite{tak}  (and by Baryen and Flato independently \cite{BF})
as a natural condition for his definition of a Nambu-Poisson manifold
which allows him to formulate the Nambu mechanics in an invariant
geometric form similar to that of Hamiltonian mechanics.
For example, the fundamental identity implies that the time evolution
preserves the Nambu bracket. Note that however for this purpose, 
a weaker form of the fundamental identity, where two of the elements
are fixed: $k=H_1, l=H_2 $, is sufficient. 
What we have shown above is that a quantized geometry characterized by the
Nambu bracket \eq{n-3bkt} is allowed as a solution in string theory and we will
refer to the quantized geometry \eq{xxx} as quantum Nambu geometry.

\section{Analysis of the Quantum Nambu Geometry}

\subsection{Representations of the Nambu-Heisenberg 
commutation relation}

An intermediate 
question to the relation \eq{xxx} is that in what sense it characterizes 
a  new quantized geometry.  We will address this question in this section.

\subsubsection{Finite dimensional Lie algebraic representations }

Let us start with the observation of Nambu \cite{nambu} 
that if  $X^i = \a l_i$ for a constant $\a$ and $l_i$
are the generators of the standard $SU(2)$ algebra
\be
[l_i, l_j ] = i \e_{ijk} l_k, 
\ee
then 
\be
[X^i,X^j,X^k] = i  \e^{ijk} \a^3 C_{\rm R}, 
\ee
where $C_{\rm R}$ is the quadratic Casmir 
for the representation ${\rm R}$ where $X^i$ is in. 
For $N\times N$ matrices, $C_N = (N^2-1)/4$ and so if we choose 
$\a^3 = \th/C_N$, 
then we can realize the relation \eq{xxx} with $N\times N$
matrices. Nambu has also constructed a representation of the relation \eq{xxx}
in terms of $SU(2)\times SU(2)$ representations. In these representations Nambu 
constructed, the quantum Nambu bracket is embedded in  
an underlying Lie algebra ($SU(2)$ or
$SU(2) \times SU(2)$) as a Casmir. As such, the relation \eq{xxx} is not 
fundamental but is a result of an underlying Lie algebraic structure.
This is pretty much the story for finite $N$.
What we will show next is that in the large $N$ limit, there are new 
representations of \eq{xxx} that are not of the above form, i.e. not
representations of any Lie algebra. It is the existence of these
representations that demonstrates the fundamental and novel nature of the 
Nambu-Heisenberg commutation relation \eq{xxx}.

\subsubsection{Infinite dimensional representations}

An infinite dimensional representation of \eq{xxx} has been 
constructed by Takhtajan \cite{tak}. However his representation is 
complex as the 
operators $X^i$ are not represented as Hermitian operators there. 
As a  result, the quantum space is six dimensional. 
In this subsection,  we give two examples of representations where 
a unitary condition can be imposed and the quantum space is three dimensional.
We remark that in the large $N$ limit, 
there is probably an infinite number of inequivalent representations for the 
operator relation \eq{xxx}. Precisely which representation is to 
be used is a 
question that depends on the physics under consideration.

To be concrete, we are interested in constructing representations of 
the relation
\be \label{x123}
[X^1,X^2,X^3] = i \th, 
\ee
where $\th$ is real and there is a certain reality condition 
which one can impose
so that the quantum space \eq{x123} can be 
understood as a deformation of a real 
3-dimensional space.

\vskip 0.3cm
\noindent \underline{1. A representation in terms of $Z,\Zb,X$}

Let us consider Hermitian $X^i$'s and introduce the complex coordinates
\be
Z:= X^1+ i X^2, \quad \Zb := X^1 -i X^2.
\ee
The relation \eq{x123} can be written in the form
\be \label{xzz}
[X, Z, \Zb] =2 \th,
\ee
where $X=X^3$.
We consider an ansatz for a representation
\begin{subnumcases}{} 
Z\ket{\o}=f_1(\o)\ket{\o+\b}+f_2(\o)\ket{\o-\b}, \label{repa}\\
\Zb\ket{\o}=f_2^*(\o+\b)\ket{\o+\b}+f_1^*(\o-\b)\ket{\o-\b},\label{repb}\\
X\ket{\o}=g(\o)\ket{\o}, \label{repc}
\end{subnumcases}
where the state $\ket{\o}$ is parameterized by a number $\o$ and 
$\b$ is a fixed ``step''.
It is clear the domain of $\o$ is one-dimensional. Without loss
of generality we can take $\b$ real and $\o \in \RR$.
The form of \eq{repb} is fixed by \eq{repa} by requiring
$\Zb = Z^\dagger$.  Hermiticity of $X$ requires that $g$ be real.
We remark that 
the introduction of $Z, \Zb$ is motivated by the creation and annihilation 
operators for the Heisenberg commutation relation. Thus it 
would be natural to 
consider the representation \eq{repa}-\eq{repc} with $f_2=0$ or
$f_1=0$. 
However this 
always give a constraint of the form
$Z \Zb + \Zb Z = \cZ(X)$ for some function $\cZ$ and so describes at most a 
2-dimensional space. As a result, we are prompted to 
try the more general ansatz 
stated above. 

It is easy to obtain
\be
[X,Z,\Zb]\ket{\o}=I_2(\o)\ket{\o+2\b}+I_{-2}(\o)\ket{\o-2\b}+I_0(\o)\ket{\o},
\ee
where
\bea 
I_2(\o)&=& G(\o) K(\o), \qquad I_{-2}(\o)= I_2(\o-2\b)^*, \\
I_0(\o) &=&  F(\o) ( 2g(\o)-g(\o-\b))  -F(\o+\b) ( 2g(\o)-g(\o+\b))
\eea
and
\bea
G(\o)&:=& g(\o+2\b)+g(\o)-g(\o+\b),\\
K(\o)&:=&f_2(\o+\b)^*f_1(\o+\b)-f_1(\o)f_2(\o+2\b)^*, \\
F(\o)&:=&|f_1(\o-\b)|^2-|f_2(\o)|^2.
\eea
We would like to find functions $g, f_1, f_2$ such that
\be
I_2= I_{-2} =0
\ee
and 
\be \label{i0}
I_0 = 2\th. 
\ee
The first condition can be solved by  requiring 
$K(\o)=0$ or $G(\o) =0$.
The possibility of $K=0$ is not good since it implies that 
$[Z, \Zb] \ket{\o} = F(\o) \ket{\o}$ and so there is a relation of the form 
$[Z,\Zb] = \cZ(X)$ for some function $\cZ$. This means the relation \eq{xzz}
is not intrinsic but reducible to a statement about commutators, this is
not we are after. For this
reason, we consider the second possibility
\be \label{G0}
g(\o+2\b)+g(\o)-g(\o+\b)=0.
\ee 
It is easy to see that it
implies a  pseduo-periodic condition
\be \label{p-g}
g(\o+3\b)=-g(\o),
\ee
and it follows that
\be
I_0(\o) =  F(\o) A(\o) -F(\o+\b) A(\o-\b),  
\ee
where
\be
A(\o) := g(\o)+g(\o+ \b).
\ee
It is 
$ A(\o+3\b) = -A(\o)$, $F(\o+3 \b) = - F(\o)$.
The condition \eq{G0} is solved by
\be \label{g-general}
g (\o) = \sin \a \o, \quad \cos \a \o, \qquad \mbox{where} \quad 
\a = \frac{\pi}{3 \b}(6 p \pm 1), \quad p \in \ZZ,
\ee
or generally a Fourier sum of these modes. For simplicity, let us 
construct a representation for the simple mode
\be
g (\o) = \cos \a \o, 
\ee
where $\a$ is as specified in \eq{g-general}. 
Consider the ansatz
\be
F(\o) = k \sin ( \a \o- \frac{\a\b}{2}) .
\ee
One sees that \eq{i0} is solved with
\be
k = - \frac{2 \th}{\sin{\a \b}  \cos\frac{\a \b}{2}}.
\ee
This provides a constraint on the two functions $f_1$ and $f_2$. For example,
a simple solution is
\be
|f_1(\o)|^2 = |f_2(\o)|^2 = k_0 - \frac{8\th}{3} \cos \a \o, 
\ee
where $k_0 > 8\th/3$ 
is  any constant such that the right hand side above is positive. Without loss 
of generality, we can take $\b=1$. The representation space is 
given by the 1-dimensional lattice
\be
\{ \ket{ \o + n}: n \in \ZZ  \}
\ee
and is of countably infinite dimension 
for each fixed $\o$.

\vskip 0.3cm
\noindent \underline{2. A representation with $Z_3$ symmetry }

We now demonstrate that there is another way to construct a representation of 
\eq{xxx} such that the quantum space it represents is 
3-dimensional. 
In this construction, we assume no reality condition on the fields $X^i$, 
so thus far we have 6 degrees of freedom. 
Instead, let us introduce a unitary operator, 
\bea
U\ket{\o} = &&\ket{\r^2 \o}, \nn \\
U^\dagger \ket{\o} = &&\ket{\r\o},
\eea
and assuming 
\be
X^1\ket{\o} = (\o +a)\ket{\o +1},
\ee
one obtains 
\bea
U^\dagger X^1 U \ket{\o} = &&(\r^2\o +a)\ket{\o +\r}, \nn \\
U^\dagger{}^2 X^1 U^2 \ket{\o} = &&(\r\o +a)\ket{\o +\r^2},
\eea
where $a \in \mathbb{C}$ and $\r$ 
is a cubic root of unity ($\r^3=1$) which is not equal to 1.

Now if the fields $X^1$, $X^2$ and $X^3$ are unitarily related to each other by
\bea
X^2 = &&U^\dagger X^1 U, \nn\\
X^3 = &&U^\dagger X^2 U, 
\eea
then
\bea
\label{evals1}
X^1\ket{\o} && = (\o + a)\ket{\o +1}, \nn\\
X^2\ket{\o} && = \r^2(\o + a\r)\ket{\o +\r}, \nn\\
X^3\ket{\o} && = \r(\o + a\r^2)\ket{\o +\r^2}
\eea
and it easy to see that 
\be
[X^1,X^2,X^3]\ket{\o} = 3(a^2-a)(\r-\r^2)\ket{\o},
\ee
where $a \in \mathbb{C}$ and $\r-\r^2$ is pure imaginary. 
In this representation the fields $X^1,X^2,X^3$ are not Hermitian. 
They are however related through a unitary transformation,
$U = e^{i\Th}$, where $\Th$ is some Hermitian operator. 
So in this representation, we have 2 degrees of freedom from $X^1$ 
and one from $\Th$ giving us 3 real dimensions.

We note that in this representation, 
the operators $X^i$'s can be constructed as pseudo-differential 
operators acting
on functions $\ipr{\o}{\psi}=\psi(\o).$ Let us start with $X^1$ and note that
$ \bra{\o~+~1}X^1=(\o+a)\bra{\o}$ and so
$ X^1\psi(\o)=\bra{\o}X^1\ket{\psi}=(\o+a-1)\psi(\o-1)$. 
Therefore, we obtain 
\be
X^1=(\o+a-1)e^{-\frac{\pa}{\pa \o}}.
\ee
Similarly
\be
X^2=(\r^2\o+a -1)e^{-\r\frac{\pa}{\pa \o}},
\ee
\be
X^3=(\r\o+a-1)e^{-\r^2 \frac{\pa}{\pa \o}},
\ee
and for the unitary operator
\be\label{eq:UnitaryPseudo}
U=\exp\left[\ln(\r)\o\frac{\pa}{\pa\o}\right].
\ee
The Hermitian conjugate of the unitary operator is
\be
U^\dagger=\exp\left[\ln(\r^2)\o\frac{\pa}{\pa\o}\right].
\ee

In this construction, the representation space is 
given by  the 2-dimensional lattice
\be
\{ \ket{ m+ n \rho }: m,n \in \ZZ  \}
\ee
and is of countably infinite dimension.

In conclusion, we have shown that 
there are at least two ways to represent \eq{xxx} as a three-dimensional 
quantum space: either a real representation, or having one complex field 
and introducing a unitary operator relating $X^2,X^3$. 
This is in contrast to the representation in \cite{tak} where all the fields 
are complex and not unitarily related.

\subsection{Integrals}

It is an interesting question to construct quantum field theory on 
the quantum Nambu geometry. An important ingredient that is needed 
is an invariant
integral on the space. Given a general quantum space, sometimes the symmetry
is strong enough to determine the integral purely algebraically. For example, 
this is the case for a compact Lie group and some homogeneous spaces of quantum 
groups. For the Nambu geometry, this is not the case due to the existence 
of  many inequivalent representations, an integral must be defined using 
information beyond the algebraic commutation relations. 
With  the representations available, 
we can use the trace to define an integral. The 
properties of the integrals as well as the construction of the quantum
field theory will be reported elsewhere.

\section{Matrix Theories in Large RR  Flux Background}
 
We would like to perform an expansion around the quantum Nambu
geometry and ask what kind of gauge theory would come out. We recall that
for the
standard Lie algebraic type noncommutative geometry, a 2-form field
strength 
is obtained from the fluctuation over the noncommutative geometry
as
\be
F^{\m\n} = -i [X^\m,X^\n].
\ee 
For
our quantum Nambu geometry, it is suggestive to interpret the quantum
Nambu bracket of the target space coordinate fields $X^\m$ 
as a 3-form field strength
\be 
H^{\m\n\l} = -i [X^\m,X^\n, X^\l]
\ee
and we would like to check this idea.

Places where a non-abelian 3-form field strength lives are, for example, 
multiple D4-branes (where the 3-form field strength would be 
the Hodge dual to a 2-form field strength) 
and multiple M5-branes (where the 3-form
field strength would be self-dual). To check the idea, we would like to
connect to these systems from our D1-branes system. 
And to do this, let us 
first
derive the Matrix model descriptions for the \IIB string theory,
M-theory and
\IIA string theory  in a large  flux background using our 
description \eq{S-C2} for the D1-branes.

\subsection{\IIB Matrix Theory }
 
The \IIB matrix model 
can be obtained by a large $N$ reduction \cite{EK} of the 
D1-string action. 
Let us first denote the
covariant derivative  $i D^\a = i \del^\a + A^\a$, $\a=0,1$ as 
\be
 i D^\a  = X^\a,
\ee
and rewrite $L_1$, $L_2$ in terms of the $X$'s:
\be\label{L1}
L_1 = - \frac{1}{2} \Tr X^i [X^\a,X^j] [X^\b,X^k] \e_{\a\b} \e_{ijk},  
\ee
\be\label{L2}
L_2 =-\Tr[X^0,X^1][X^2,X^3,X^4].
\ee
Although the form of $L_1$ does not look like it, it is not hard to
show that 
\be
L_1+ L_2 =\frac{1}{40} \Tr[X^a,X^b][X^c,X^d,X^e]\e_{abcde}.
\ee 
It is quite remarkable that the D1-branes Chern-Simons coupling to a
constant RR $F_3$ flux can be
written in such a simple form. 
The action of $N$ D1-branes in a large $F_3$ limit can thus
be written compactly as
\be
\label{D1}
S_{D1} = \frac{\m_1 f}{40}\int  
d^2\s\,\Tr[X^a,X^b][X^c,X^d,X^e]\e_{abcde}
= \frac{3 \m_1 f}{10 }\int  d^2\s\,\Tr X^a X^b X^c X^d X^e\e_{abcde},
\ee
where $a,b,c,d,e=0,1,2,3,4.$ 
The large $N$ reduction gives immediately 
the following D-instantonic action 
(ignoring an  unimportant overall numerical constant),
\be
\label{D-1}
S_{IIB} = 
\frac{f}{g_s l_s^2} \Tr X^a X^b X^c X^d X^e\e_{abcde}, \quad
a,b,c,d,e=0,1,2,3,4. 
\ee
This gives the matrix model description for the \IIB string theory
in the limit of a large constant RR 3-form flux, and in the
sector with  $X^{a'} =0$, $a' = 5,6,7,8,9$.
In this limit, the Myers term dominates 
over the standard Yang-Mills term in the IKKT matrix model.

\subsection{Matrix model of M-theory}

The \IIB background we considered is invariant under the Killing
vector $\del/\del x^i, i =2,3,4$. Therefore we can compactify, say
$x^2$ on a circle of radius $R_2$ and T-dualize. The corresponding \IIA
background has:
\bea
\mbox{metric}:&& S^1 \times \RR^2 \times AdS_2 \times S^5, 
\label{IIA-bkgd-1}\\
\mbox{constant RR field strength}: && \nn\\
F_{ij} &=& F_{2ij}, \quad i,j =3,4, \nn \\
F_{abcd} &=& F_{2abcd}, \quad a,b,c,d = 0,1,3,4, \label{IIA-bkgd-2}\\
F_{2a'b'c'd'e'} &=& F_{a'b'c'd'e'}, \quad a',b',c',d',e'= 5,6,7,8,9, \nn\\
\mbox{constant dilaton:} && 
e^{\phi'} = e^\phi \frac{\sqrt{\a'}}{R_2}. \label{IIA-bkgd-3}
\eea

Under T-duality, the D1-branes become D2-branes. In the double scaling
limit \eq{scaling}, the D2-branes action is given by 
the T-dual of the D1-branes
action \eq{D1} by applying  the usual T-duality rule
\cite{bfss, taylor} to the D1-branes action:
\bea
X^2 = i R_2 D_2, \label{Td1} \\
\Tr = \int \frac{l_s d \s_2}{R_2} \Tr. \label{Td2}
\eea 
We obtain 
\be \label{D2}
S_{D2} = \frac{f}{g_s l_s } \int d^3 \s X^a X^b X^c X^d X^e\e_{abcde}
\ee
where $a,b,c,d,e =0,1,2,3,4$ 
and we have ignored an  unimportant overall numerical constant.
Note that since the Chern-Simons coupling is topological,  
the $R_2$ dependence gets cancelled in \eq{D2}. 
We note that one can also obtain \eq{D2} directly from the 
Chern-Simons coupling of
D2-branes in the \IIA RR flux background \eq{IIA-bkgd-2}. It is
\be
S_{CS} = \frac{1}{g_s l_s^3} \left[
\int P(C_1) F \l + \int P(C_3) + 
\int P(i \l \ri_\Phi \ri_\Phi C_5)   
\right].
\ee
The $C_3$ and $C_5$ terms have their origin from  the RR
5-form of \IIB and so they can be ignored in the 
double scaling limit \eq{scaling}. The $C_1$ term then
reproduces precisely \eq{D2}. 

In addition to D2-branes, the \IIA side also contains D0, D4, D6 and 
D8-branes. If we put M-theory on a circle and go to the infinite
momentum frame, then only states with positive D0-brane charge are
left in the physical description. 
In general, this includes all the
D-branes in the \IIA theory since by turning on a worldvolume
Born-Infeld configuration $F\wedge \cdots \wedge F$ with $p/2$ terms, 
a D$p$-brane is charged under $C_1$. 

With remarkable insights, BFSS  proposed originally that M-theory in flat space 
in the infinite momentum frame  is given by the large $N$ quantum mechanics of
D0-branes \cite{bfss}. 
The reason why it is not necessary to include the higher 
D$p$-branes ($p$ even) 
is because they can be constructed out of the D0-branes and so they 
are already included. This is so is because in flat space, 
the worldvolume action for such a system of D$p$-branes is 
given by 
\be
S_{YM}  = \int d^{p+1} \s \; [X^I,X^J]^2,
\ee
where $X^I = (X^\m, X^i)$, $\m =0, 1, \cdots, p$, $i =p+1, \cdots, 9$
and $X^\m = i D^\m$ and a background with nontrivial $F^{\wedge n}$ is assumed.
In this way one can see that all the higher
D$p$-brane actions can actually be constructed from the D0-branes and so it is
sufficient to include only the D0-branes in the description. 

For us, we would like to derive the quantum mechanical 
description of M-theory in a curved background that corresponds to
\eq{IIA-bkgd-1} - \eq{IIA-bkgd-3} uplifted to 11 dimensions.
The eleven dimensional background reads
\bea
\mbox{metric}:&& ds^2 = e^{-\frac{2}{3} \phi} ds^2_{IIA} +
 e^{\frac{4}{3} \phi}  (dx^{11} - dx^i C_i)^2,
\label{M-bkgd-1}\\
\mbox{3-form potential}: && C^{(3)} = \frac{1}{6} C_{abc} dx^a dx^b dx^c, 
\label{M-bkgd-2}
\eea
where $\phi$ is the dilaton in \IIA theory, $C_i$ and $C_{abc}$ are the RR
1-form potential and RR 3-form potential which appear 
in \eq{IIA-bkgd-2}.

Let us denote the 11-th dimensional radius by $R_{11}$. 
In general, with a suitable worldvolume flux turned on, 
the higher D$p$-branes ($p$ even) carries D0-brane charges and so in principle 
should be kept in the infinite momentum frame. However as in the flat case,
it is sufficient to select a
subset of degrees of freedom in such a way that 
all the other degrees of freedom as well as their dynamics could be
recovered. 
Now what is different for our background  is that there is a set of
non-vanishing RR gauge potentials which lead to explicit
Chern-Simons terms in the action of the D$p$-branes.

Let us examine this in detail. 
In the double scaling limit \eq{scaling}, 
we can ignore the Yang-Mills term and the Chern-Simons coupling to
$C_3$ and $C_5$ (whose origin are both from $F_5$ of the
\IIB side) and 
concentrate on the Chern-Simons coupling of $C_1$. 
Moreover in the sector where
the fields in the sphere directions are set to zero: $X^{a'} =0$, $a'
= 5,6,7,8,9$, the Chern-Simons couplings for D4, D6 and D8-branes are
zero. For the D0-branes, we have 
(ignoring an  unimportant overall numerical constant), 
\be\label{D0}
S_{D0} = \frac{1}{g_s l_s} \int P(C^{(1)}) 
=  \frac{f}{g_s l_s} \int dt \e_{ij} X^i D_t X^j , \quad
i,j =3,4.
\ee 
Now  the action \eq{D2} is equivalent to it's
dimensional reduction
\be \label{red}
\frac{f}{g_s l_s} \int  dt \,\Tr X^a X^b X^c X^d X^e\e_{abcde},
\ee
since one can always recover $S_{D2}$ by compactifying $X^1, X^2$ 
and then decompactify using the rules \eq{Td1}, \eq{Td2}. 
Since the action \eq{D0} can be
considered as a special case of 
\eq{red}
in a background $[X^1,X^2] = 1$,
therefore  
we propose that in the large flux limit and in the
sector with  $X^{a'} =0$, $a' = 5,6,7,8,9$, M-theory in our curved background 
\eq{M-bkgd-1}, \eq{M-bkgd-2} 
is described by the quantum mechanical action 
\be
\label{M}
S_M 
= \frac{if}{g_s l_s } \int dt \,\Tr D_t X^b X^c X^d X^e\e_{bcde},\quad
b,c,d,e=1,2,3,4.
\ee
Here we have substituted $X^0 = -i D_t$ and we have 
ignored an  unimportant overall numerical constant.

\subsection{\IIA Matrix String Theory}

Given the Matrix model \eq{M} for M-theory, 
one could follow the procedure of \cite{dvv} and derive
the corresponding \IIA Matrix string theory. To do this, we first
rewrite \eq{M} in terms of  the eleventh dimensional radius
\be
R_{11} = g_s l_s,
\ee
then compactify $X^2$ on a circle of radius $R_2$, 
and finally perform an 11-2 flip which
exchanges the role of the 11th and the 2nd
direction of the $T^2$ where our M-theory is compactified on.
In practice this amounts to having instead
\be
R_2 = g_s l_s, 
\ee
and 
\be
R_{11} =N,
\ee
where a normalization of lightcone momentum $p_+ =1$ is adopted
\cite{dvv}. In this way, we obtain the Matrix string description
\be
\label{IIA}
S_{IIA}
= \frac{f}{N}
\int  d^2 \s \,\Tr X^a X^b X^c X^d X^e\e_{abcde}, \quad
a,b,c,d,e=0,1,2,3,4, 
\ee
where 
\be
X^\a = i D^\a, \quad X^i = \mbox{scalars}, \qquad \a= 0,1, \quad i = 2,3,4
\ee
and we have ignored an  unimportant overall numerical constant.
We note that the D1-strings action \eq{D1} 
and the \IIA Matrix string action \eq{IIA} are indeed the same up to a constant
coefficient. This is similar to what was found in \cite{dvv,ver,bonora} 
where the same 
2-dimensional supersymmetric Yang-Mils theory could have different string
interpretations depending on how one associate its parameters with the string 
theories. This is consistent with T-duality.

\section{Multiple D4-Branes and M5-Branes}

\subsection{D4-branes in large RR 2-form flux}

Let us concentrate on the Matrix string theory.
Since the Matrix string theory \eq{IIA} takes the same form as the
original D1-strings action \eq{D1}, it follows immediately
that it admits the classical solution:
\be \label{soln-23}
[X^\a, X^\b] =0, \quad [X^\a, X^i ] =0, \quad \a=0,1,\; i =2,3,4.
\ee
As before, the commutation relations of $X^i$ among themselves are 
not constrained. Let us consider the solution  
$X^i_{cl} =x^i$ of quantum Nambu geometry
\be \label{d4-xxx}
[x^2,x^3,x^4] = i\th
\ee
and 
consider a fluctuation around it. 
In the large $N$ limit, we get a set of large $N$ matrices $x^i$.
Depending on
the representation chosen, they may or may not
generate the entire set of $N\times N$ matrices. In general, assume $x^i$
do not generate the whole set of $N \times N$ matrices. Then every $N\times N$
matrix can be expressed as a $K\times K$ matrix whose entries are functions of
$x^i$ \cite{bkgd-indep}.
The expansion of the dynamical variables around the classical solution
can thus be parameterized as
\be
X^i = x^i \id_{K \times K} + A^i(\s,x^j).
\ee
The action \eq{IIA}  becomes
\be \label{S5}
S_5 =  \frac{f}{N} \int_{\S_5} \tr X^a X^b X^c X^d X^e\e_{abcde}
\ee
where   $\int_{\S_5} = \int d^2 \s \int_x$  and $\int_x$ is an integral on
the quantum Nambu geometry which can be constructed from a representation of
the geometry. In the large $N$ limit, the trace over large $N$ 
matrices decompose
as usual as $\Tr = \int_x \tr$.

We would like to argue 
that this solution corresponds to a system of $K$ parallel 
D4-branes.
To do this, let us 
introduce a three-form $H$-field whose components are defined
by
\bea 
H^{abc} &= & - i [X^a,X^b,X^c], \label{H-id-1} \\
H^{de 5} &= & - i  [X^d, X^e], \qquad a,b,c,d,e = 0,1,2,3,4,
\label{H-id-2}
\eea
where 
\be
 H^{*\m\n\l} := \frac{1}{6 \sqrt{-g}} \e^{\m\n\l\r\a\b}H_{\r\a\b}
\ee
is the Hodge dual of $H_{\r\a\b}$.
Our convention for the  Hodge duality operation is
$\e_{012345} =1 =-g \e^{012345}$. 
We remark that a similar
identification has also been proposed in
\cite{CS1} in the analysis of the M5-brane geometry in a large $C$-field.
As a result we obtain
\be \label{SHH}
S_5 =  \int_{\S_5} \tr H^{abc} H^{de5} \; \e_{abcde},
\ee
where have ignored an unimportant overall constant here.

To see the connection of \eq{SHH}  with D4-branes, let us 
consider the abelian case. 
Based on important earlier works \cite{schw}, 
PST constructed a covariant action for 
for a self-dual 3-form field strength  $H= dB$  
living on a  single M5-brane \cite{pst}:
\be \label{pst}
S_{PST} = \int d^6 \s  \left[\sqrt{-g} 
\frac{1 }{4(\del a)^2}\del_{\m} a
H^{*\m \n \l} H_{\n \l \r} \del^\r a + Q(\Ht)
\right].
\ee
Here the Greek indices $\m=0,1,\cdots,5 $ and
\be
\Ht_{\m\n} := \frac{1}{\sqrt{(\del a)^2}}H^*_{\m\n\l} \del^\l a.
\ee
The action \eq{pst} is invariant under the following local transformations:
\bea  
(I) && \quad \d B_{\m\n} = \del_{[\m} \L_{\n]}, \quad \d a =0;\label{T1}  \nn\\
(II) &&\quad \d B_{\m\n} = \del_{[\m } a \; \vphi_{\n]}
\quad \d a =0; \label{T2}\nn\\
(III)&& \quad \d B_{\m\n} 
= \frac{\vphi}{2 (\del a)^2} (H_{\m\n\r} \del^\r a - \cV_{\m\n}),
\quad \d a = \vphi, \label{T3}
\eea
where
\be
\cV^{\m\n} := -2 \sqrt{\frac{(\del a)^2}{-g}} \frac{\d Q}{\d \Ht_{\m\n}}.
\ee
The equation of motion of the 2-form potential $B_{\m\n}$ is
\be\label{pst-eom}
\e^{\m\n\l\r\a\b} \del_\l \left(
\frac{\del_\r a}{(\del a)^2} (H_{\a\b \g} \del^\g a - \cV_{\a\b})
\right) =0.
\ee
Using the local symmetry \eq{T2}, one can then show that it is equivalent to
the self-duality condition 
\be \label{sdeom}
H_{\m\n\l} \del^\l a - \cV_{\m\n} =0. 
\ee
The scalar field $a$ is introduced to allow six dimensional covariance
and is completely auxiliary due to the symmetry \eq{T3}. 
If we choose a gauge $a=x^5$ and consider the linearized case (i.e. linearized 
equation of motion) with
\be
Q= -\frac{1}{4} \Ht_{\m\n} \Ht^{\m\n} \sqrt{-g},
\ee 
then 
\be 
\cV_{\m\n} = \Ht_{\m\n} \sqrt{(\del a)^2}
\ee
and \eq{sdeom} becomes the standard self-duality condition
\be
H_{\m\n 5} = H^*_{\m\n 5}.
\ee
In this case the gauge fixed PST action reads
\be \label{pst1}
S_{PST} = -\frac{1}{4}\times  \int d^6 \s\left(
 \frac{1}{6} \e_{abcde}H^{abc} H^{de5} + H^{* ab5} H^*_{ab5}  \sqrt{-g} 
\right),
\ee
where $a=0, \cdots, 4$ etc. See \cite{schw2} 
for a detailed discussion of the non-covariant and covariant PST
formulations of the M5-brane action. Another equivalent description is 
to use the superembedding approach \cite{sm,equiv}.
In the next subsection,
we will argue how to generalize the PST description for a single
M5-brane  to the non-abelian case.

Now if we perform a dimensional reduction on $x^5$, the 
first term $\e_{abcde} H^{abc} H^{de5}$  in the dimensionally
reduced gauge fixed PST action  is  precisely equal to \eq{SHH}.
This matching is quite amazing.  
As for the second term $H^{*ab5} H^*_{ab5}  $, it is identified with the
D4-branes' Yang-Mills Lagrangian  $\sqrt{-g} F_{ab}^2$ by performing a 
(Hodge) dualization. 
However we have shown above that the Yang-Mills term is 
negligible in our double scaling limit, therefore the  $H^{*ab5} H^*_{ab5}  $ 
term is  not seen in the action 
\eq{SHH}. 
Since a dimensionally reduced M5-brane is simply a D4-brane, 
this means \eq{SHH},
for the Abelian case, 
does describe a D4-brane in the large RR flux background. 
For the non-abelian case, we  propose  that the action \eq{SHH}
describes a sector of multiple D4-branes theory  (where $X^{a'} =0$) 
in a large RR 2-form flux background.

We note that since our term \eq{SHH} agrees with the PST action \eq{pst1} 
only if the map
\eq{H-id-1} is employed, the matching gives us confidence in this
identification.
We emphasise that the reason that it is possible to write \eq{S5} in 
terms of the $H$'s is entirely due to the fact that the
D1-branes Chern-Simons action could be 
combined nicely into the remarkable form  \eq{D1},  which
is true only for our constant RR-flux 
in the \IIB background.

\subsection{A proposal for a 
theory of multiple M5-branes using 1-form gauge field} 

Recently, it has been argued that \cite{d1,d2} 
the instantons on multiple D4-branes could be 
identified with the KK modes associated with the compactification of M5-branes 
on a circle. By including all these modes, it was proposed that 
the low energy SYM theory of D4-branes is 
a well-defined quantum theory and is actually the theory of multiple
M5-branes compactified on a circle. Back to our proposed action 
\eq{SHH} for D4-branes in a 
large RR flux background, how can we incorporate the higher KK modes in our 
description? 
A possible
hint is from the identification \eq{H-id-2}. We note that the
identification  for
$H^{de5}$ can be written as
\be \label{H-id-3}
H^{de5} =  -i [X^d,X^e,X^5]
\ee
 with 
\be \label{X5-1}
X^5 = \id.
\ee
If we think about $X^5$ as a scalar field describing the compactified 
$X^5$ direction transverse to the D4-brane, then one can understand
the relation \eq{H-id-2} and \eq{X5-1} as saying only the zero mode of
the M5-branes has been included, i.e. a dimensional reduction to D4-branes.   
In this picture, it is suggestive to include the higher KK/instantonic modes 
by promoting $X^5 =\id$ to a general field.
The identification \eq{H-id-1} and \eq{H-id-3} can be put together as
\be \label{H-id-4}
H^{ \m\n\l} = -i [X^\m, X^\n,X^\l].
\ee
We would
like to propose that 
it is a different way to
write the non-abelian self-dual 3-form field strength living on M5-branes.
In a conventional description, there would be a  non-abelian 2-form potential
$B$ and $H =dB + \cdots$ where the $\cdots$ term denotes terms necessarily for
the non-abelianization. Thus we are proposing that there is a dual
description 
of 
the non-abelian 3-form field strength in terms of the 1-form variables 
$X=X_\m d\s^\m$; 
and the $B$-field and the $X$-field are related, although one 
can expect the relation to be very complicated. 
To justify our proposal, 
one needs
to show that $H^{\m\n\l}$  satisfies the correct equation of motion 
(i.e. the self-dual equation) and describes the correct number of 
on-shell degrees of freedom (i.e. three). We will now construct 
an appropriate action to try to achieve these goals.  

Let us start with the following action
\be \label{SM5-six}
S_{M5} = -\frac{1}{4} \int_{\S_6} \tr
\left(
 \frac{1}{6} \e_{abcde}H^{abc} H^{de5} + {\big(} 
c_2 H^{abc} H_{abc}
+ c_3  H^{ab5} H_{ab5}
 {\big)} \sqrt{-g}
\right),
\ee
where $\S_6 = \S_5 \times S^1$ is the worldvolume of the M5-branes and
\eq{H-id-4} is to be used. We will consider a constant metric.
The action \eq{SM5-six} is the most general quadratic 
action that can be constructed out of the components 
$H^{abc}$ and $H^{ab5}$ and which is compatible with the $SO(1,4)$ Lorentz
symmetry. For a non-abelian generalization of the PST Lagrangian \eq{pst1}, 
it is expected that $c_2=1$ and $c_3 =0$. Here
we have allowed for more general possibility since there is no reason to
expect that our action to be exactly the same.

Our goal is to construct an action for the non-abelian 3-form field 
strength living on a system of M5-branes.
Generally, one can turn
on a constant $C$-field on the worldvolume of the M5-branes. 
How could one incorporate a $C$-field in 
\eq{SM5-six}? It is useful to recall a similar 
story for the case of 
D-branes where it is well known that a constant NSNS $B$-field can be naturally 
included as
classical solution (corresponds to a non-commutative geometry) 
of  matrix models \cite{bss,bkgd-indep,makeenko,smith-2b}. The 
remarkable feature
of this construction is that 
the different backgrounds that correspond to different $B$-fields arise as 
different classical solutions of the same degrees of freedom of the
underlying matrix model. 
Therefore let us follow the same route and  consider a reduction of
the matrix model  to a point. As a result, we get the matrix model 
\be \label{S0}
S_0 =   \frac{1}{4} \Tr \left(
c_1 \e_{abcde} X^a X^b X^c X^d X^e X^5 
+ c_2 [X^a,X^b,X^c]^2
+ c_3   [X^a,X^b,X^5]^2
\right),
\ee
where
\be
\label{cc}
c_1 = 2
\ee 
and the parameters $c_2, c_3$ are to be determined.
We will now require that the equation of motion of the matrix model to agree
with the self-duality condition of $H$. Quite remarkably this can be 
achieved with a particular choice of the parameters. 
 
The equations of motion of $S_0$ are
\be \label{eom1}
c_1 \e_{abcde} X^a X^b X^c X^d X^e +2 c_3 [[X_a,X_b,X_5], X^a, X^b]'=0, 
\ee
and
\bea \label{eom2}
&& c_1  \e_{abcde} 
\left( X^b X^c X^d X^e X^5 + X^b X^c X^d X^5 X^e +X^b X^c X^5 X^d
  X^e +X^b X^5 X^c X^d X^e \right. \nn\\
&&\qquad \qquad \left. + X^5 X^b X^c X^d X^e \right) \nn\\
& +& 6 c_2 [[X_a,X_b,X_c],X^b,X^c]' 
+ 4 c_3 [X_a,X_b,X_5],X^b,X^5]' 
=0.
\eea
The first equation \eq{eom1} can be written as
\be
(\frac{c_1}{6} -c_3) \e_{abcde} X^a X^b H^{cde} 
+ 2c_3[H_{ab5} +\frac{1}{6} \e_{abcde}H^{cde}, X^a,X^b]' =0.
\ee
Since we want to interpret $H^{\m\n\l}$ of \eq{H-id-4}  as the 
non-abelian field strength on M5-branes, $H^{\m\n\l}$ must satisfy a 
Bianchi identity. The most natural gauge covariant version would be
\be \label{bianchi}
[X^{[\m}, H^{\n\l\rho]}]=0.
\ee
We use a convention of $[X^{[a}, H^{bcd]}] = [X^a, H^{bcd}] -  [X^b, H^{cda}]+  
[X^c, H^{dab}] - [X^d, H^{abc}]$.
We will comment on its possible origin later.
Let us assume this condition holds, particularly 
\be \label{bi1}
[X^{[a}, H^{bcd]}] =0,
\ee
then
we see that the self-duality condition
\be \label{sd1}
H_{ab5} = -\frac{1}{6} \e_{abcde} H^{cde}
\ee
solve \eq{eom1}. 

Next we turn to  \eq{eom2}. Using
the conditions \eq{bi1} and the 
self-duality condition \eq{sd1}, one can show that
the LHS of \eq{eom2} can be written as
\be \label{eom2'}
(\frac{c_1}{2} + 18c_2)\{H_{ade}, X^d X^e \} 
+\{\frac{1}{4} B^{bcd5} , X^e\} \e_{abcde}
+ (\frac{c_1''}{4} - \frac{2 c_3}{3}) (X^b X^5 H^{cde} - H^{cde} X^5 X^b) ,
\ee
where
\be
B^{bcd5}:= c_1'[H^{bcd},X^5] - 12 c_2 [H^{5[bc}, X^{d]}] 
\ee
and the constants $c_1', c_1''$ satisfy
\be \label{c-1}
c_1' + c_1'' = c_1.
\ee
To get this, we have 
split the term proportional to $c_1$ 
of \eq{eom2} into two terms (with coefficients $c_1'$ and $c_1''$) and
used the $c_1'$ term to combine with the $c_2$ term and the $c_1''$ term to 
combine with the $c_3$ term to arrive at \eq{eom2'}. We note that the 
term $B^{bcd5}$ is of the form of the Bianchi identity
\be\label{bi2}
[X^{[5},H^{bcd]}] = 0
\ee
if $c_1' = 4 c_2$.
Therefore the equation of motion \eq{eom2} is satisfied if the coefficients are 
such that
\be \label{c-2}
c_1' = 4 c_2, \quad c_1'' = -40 c_2, \quad c_3 = -15 c_2
\ee
and the condition \eq{bi2} is satisfied. 
 
All in all, the equations of motion \eq{eom1}, \eq{eom2} are satisfied if
the self-duality condition \eq{sd1} and the condition \eq{bianchi}
are satisfied and if the coefficients $c_i$ are given by
\be \label{c-values}
c_2=  -\frac{1}{18} (c_1/2), \quad c_3 = \frac{5}{6} (c_1/2).
\ee 
It is amazing that a set of parameters can be found consistently so that the 
self-duality condition emerges from a matrix model.
This is not guaranteed a priori and provides evidence that the matrix
model \eq{S0} has something to do with a theory of self-dual 3-form field 
strength. 

To get a six dimensional field theory, we need to
consider  classical solutions to the equations of motion and
incorporate the fluctuations around them to build the six dimensional theory.
An interesting class of solutions which are useful for this purpose is
$X^\m =x^\m$ such that
\be \label{6dnambu}
[x^\m,x^\n,x^\l] = i \th^{\m\n\l} \id,
\ee
where $\th^{\m\n\l}$ are arbitrary constants.
Clearly, the condition \eq{bianchi} is satisfied. 
Moreover the self-duality condition is satisfied if the parameter
$\th^{\m\n\l}$ is self-dual.
Thus we obtain a six dimensional quantum Nambu geometry 
parameterized by self-dual parameter $\th^{\m\n\l}$. 
The fluctuations around the solution can be written as
\be \label{flu}
X^\m = x^\m \id_{K\times K}+ A^\m(x),
\ee
where $A^\m$ are $K\times K$ matrices. The large $N$ trace becomes
\be
\Tr = \int_x \tr,
\ee
where $\int_x$ is determined from the representations of the quantum
Nambu geometry \eq{6dnambu} and our proposal for a theory of $K$ M5-branes
(or more precisely,  $K$ non-abelian 3-form) is
\be \label{SM55}
S_{M5, \th} = -\frac{1}{4} \int_x \tr
\left(
 \frac{1}{6} \e_{abcde}H^{abc} H^{de5} + {\big(} 
\a\; \frac{-1}{3} H^{abc} H_{abc}
+ (1-\a) H^{ab5} H_{ab5}
 {\big)} \sqrt{-g}
\right),
\ee
with $\a =1/6$.  We note that with 
the self-duality condition, 
the second and the third term in \eq{SM55} can be summed together 
and is equal to
$H^{ab5} H_{ab5}$ for any value of $\a$; 
and therefore 
the action \eq{SM55} has in fact precisely the same
form (including the coefficients) as the 
non-abelian generalization of \eq{pst1}.
However  only for
$\a =1/6$ can one identify a Bianchi identity and the self-duality condition.

We note that our M5-branes system has a quantum Nambu geometry 
as its worldvolume
geometry.
What is the physical origin responsible for this quantized spacetime? 
The emergence of a noncommutative worldvolume on a brane is typically the 
result of a background gauge potential being turned on in its worldvolume.
The fact that the quantization parameter $\th^{\m\n\l}$ is self-dual suggests to
identify it with the self-dual 3-form $C$-field  on the worldvolume of 
the M5-branes. This identification is further supported by the fact that if we 
dimensionally reduced the M5-branes, say, on  the 5-th direction, 
which amounts
to putting $X^5 =\id$, then the relation \eq{6dnambu} reads
\be
[X^\m,X^\n, \id] = [X^\m,X^\n] = i \th^{\m\n5}. 
\ee
This is the noncommutative geometry over D4-branes with a $B$-field whose
components are $B_{\m\n} = \th_{\m\n 5}$ 
(we remind the readers we are considering the linearized limit). 
Since the $B$-field is related to the 11-dimensional $C$-field as 
$B_{\m\n} =C_{\m\n5}$, it is correct to identify 
$\th^{\m\n\l}$ with the constant $C$-field $C^{\m\n\l}$. 
All in all, we conclude that  the geometry \eq{6dnambu} 
is the result of having a self-dual 3-form $C$-field 
\be \label{C-th}
C_{\m\n\l} = \th_{\m\n\l}
\ee
turned on in the worldvolume of the M5-branes.

In our description, the field strength $H^{\m\n\l}$ is 
constructed from the 1-form potentials $A^\m$ using \eq{H-id-4}
and \eq{flu}. 
In a conventional description of 3-form field strength, a 
2-form potential is used. 
Our analysis suggests that there maybe in fact 
two equivalent formulations for
the theory of multiple M5-branes in a $C$-field, one in terms of a 1-form
gauge field as in ours \eq{SM55}, and 
the conventional formulation in terms of a 2-form gauge potential.

Evidence of this can be seen from the 
counting of the degrees of freedom of  our model.
Initially we have six fields. If the self-duality equation is in fact
the equation of motion of the theory,
then    
the degrees of freedom  are reduced to half and 
we have indeed three degrees
of freedom which is appropriate for a description of a self-dual
3-form field strength.
The theory \eq{SM55} would then have  all the desirable
properties of a theory
of non-abelian self-dual 3-form field strength except that the theory 
is written
manifestly using a 1-form potential as the variables.

A couple of comments on the dual formulation are in order:

\begin{enumerate}

\item As noted above, our action \eq{SM55}  is equal to
the non-abelian form
of the PST action \eq{pst1}  when the self-duality condition is satisfied.
The agreement of the actions on-shell is a necessary condition for 
our formulation
to be an equivalent description on-shell. Therefore this agreement
provides
more support that our proposed action \eq{SM55} indeed provides
a dual description of the non-abelian self-dual 3-form.

\item
Our formulation of using a 1-form gauge field $A_\m$
is supposed to be equivalent to the conventional formulation of 
using a 2-form gauge potential $B_{\m\n}$ only on-shell. As such
there can be a relation between the 2-form gauge field
and the 1-form gauge field only on-shell. Identifying such a $B$-field
from our description is important as it would allow us to couple to
self-dual strings.

\item 
In the conventional formulation,  
the existence of the tensor gauge symmetry and
the self-duality equation is crucial in 
reducing the fifteen components of $B$ to three.
In our formulation, we do get the
desired 
number
of degree of freedom 
(modulo the issues discussed above) and there is no need of a tensor 
gauge symmetry.
Curiously, in a recent construction of the non-abelian 3-form theory using a 
2-form $B$-field potential \cite{chu}, it was shown that the tensor gauge 
symmetry (part of the $G \times G$ symmetry structure constructed there) could
be gauge fixed to an ordinary gauge symmetry $G$ (diagonal part of $G\times G$).
It is interesting that 
the gauge fixed theory has precisely the same gauge symmetry as our
proposed  description here. This coincidence provides 
some support to both the description proposed in \cite{chu} and the 
description proposed here.

\item  In the above we have obtained the Bianchi identity and the 
self-duality condition as a solution of
the reduced matrix description. 
However to fully justify our proposal, we need to establish
that it is the only nontrivial solution.
We recall that in the PST action \eq{pst1},
one does not get the self-duality condition \eq{sdeom}
as the equation of motion immediately.
To do this, one needs to make
crucial use of  the symmetry \eq{T2} which acts on the $B$-field.
For our case, it is possible that there is a counterpart of 
the symmetry \eq{T2} which acts on the $X$'s; and this symmetry is needed 
to derive the  self-duality equation (and hence the Bianchi identity). 
It is important to understand whether such a symmetry really exists 
in our model,
and if so, how it acts.

Another  possible way
to settle the issue is to supersymmetrize our action with (1,0) or
(2,0) supersymmetry 
since supersymmetry would require the 3-form field strength to
be self-dual automatically.  
Supersymmetrization of our system is also needed for describing M5-branes. 
In any case supersymmetry is 
an important topic and we hope to return to it in future work.
See \cite{lambert-susyM5} for some recent related works on (2,0) supersymmetry 
of a non-abelian self-dual 3-form field strength  multiplet.

\end{enumerate}

 \section{Discussions}

In this paper we have achieved the goal of finding a novel kind of quantum
geometry in string theory. The geometry we found is characterized by a 
quantum Nambu bracket and is intrinsically different from the usual 
Lie-algebraic
type noncommutative geometry. 
Starting with an analysis of the D1-branes matrix model, we arrive at
the action \eq{SM55} which we proposed to be 
the theory of non-abelian self-dual  tensor living on multiple M5-branes.
We also found that
the worldvolume of the M5-branes in a $C$-field is  described by a
quantum Nambu geometry with self-dual parameters  $\th^{\m\n\l} = C^{\m\n\l}$. 
These are  the main results of the paper.

It is intriguing that there seems to be a dual formulation of the theory 
of non-abelian  self-dual 3-form field strength in terms of a 
1-form gauge potential. We have discussed various aspects of this dual 
formulation
and how this may be related to the conventional formulation in terms 
of a 2-form
gauge field. As a proposed description for the worldvolume theory of 
multiple M5-branes, 
it is interesting to understand how \eq{SM55} could be reduced
to the non-abelian Yang-Mills theory of D4-branes. It 
is also necessary to 
include supersymmetry in \eq{SM55}.
 
It is an interesting result that the worldvolume of an 
M5-brane in a $C$-field 
background is described by a
quantum Nambu geometry with self-dual parameters. 
One may wonder how to obtain this result by
a quantization of an open M2-brane in the presence of a $C$-field. 
However it appears that  treating
the mixed boundary condition as a constraint and canonically quantizing
the system may not be the best way to proceed \cite{M2-C}. 
It may be possible that a
different choice of the quantization variables and a reformulation of the
quantization is necessary.

Our proposed action \eq{SM55} for multiple M5-branes in a $C$-field \eq{C-th} 
is defined on a quantum Nambu geometry. 
It is instructive to recall that in the case of D-branes with a $B$-field, the 
worldvolume action can be either expressed in terms of a commutative language 
as a Dirac-Born-Infeld action, or in terms of noncommutative geometry, 
as a much simpler noncommutative Yang-Mills action. This remarkable equivalence 
is established with the Seiberg-Witten map \cite{sw}. 
In our case, the action for a system of M5-branes with $C$-field can in
principle be constructed as a non-abelian generalization of the Abelian
PST action (linearized or nonlinear) with $C$-field. This action
has not been constructed (see however \cite{chu} and 
also \cite{pm,sezgin} for some recent  proposals for the case with $C=0$),
but in any case 
can be expected to be very complicated. Our proposed
action \eq{SM55} takes a much simpler form. 
Like the noncommutative Yang-Mills action, it 
is supposed to be equal to the full form of a non-abelian non-linear 
PST action with a
$C$-field via some kind of mapping like the Seiberg-Witten map. The 
understanding
of how symmetries are realized in the different models is important
in understanding this Seiberg-Witten map.

We have presented a preliminary analysis of the 
properties of the quantum Nambu geometry.
For the usual noncommutative
geometry of the Moyal type, the existence of a noncommutative
parameter leads to a number of very interesting physical effects
\cite{ncg-review}. 
It will be interesting to
construct quantum field theory on the quantized Nambu space and to uncover 
the novel physical effects associated with the existence of the
Nambu scale parameter $\th$. 
In the dual language, one can try to formulate the operatoric 
quantum Nambu geometry as an algebra of functions with a $*$-product.
The construction of the $*$-product will be very interesting and will
be very helpful 
in 
the construction of the quantum field theory.

In the usual noncommutative
geometry of the Moyal type, a fluctuation analysis typically leads to
a (noncommutative) Yang-Mills gauge theory whose field strength 
is defined by a commutator. 
We have  performed a small fluctuation analysis 
around the quantum Nambu geometry describing a system of D4-branes 
and  found that the action can be 
written in terms of a three-form field strength defined by
the quantum Nambu bracket. We have also argued for a dual formulation
of the theory of a 3-form field strength whose field strength is defined
in terms of the quantum Nambu bracket of a 1-form gauge field.
The
pattern seems to be that a
non-abelian
$N$-form gauge theory is naturally
associated with  a quantum geometry defined by a 
quantum $N$-bracket. It will be interesting to develop this further
and understand better the properties of the gauge symmetry for these 
higher-form 
gauge theories.

A close variant of the quantum Nambu geometry is the geometry
defined by the following relation
\be \label{xxx-x}
[X^i,X^j,X^k] = i \l \e^{ijkl} X^l,
\ee
which is also defined using the quantum Nambu 3-bracket. 
In the case of Lie 2-bracket, the fuzzy geometry could be obtained in a 
matrix model where the Moyal type noncommutative geometry is a solution by 
adding a mass term. 
It would be interesting to see whether
\eq{xxx-x} also arises in string theory and in particular 
from a modification of our D1-branes
matrix model with additional terms.
The geometry \eq{xxx-x} with a Lie 3-bracket has played an important 
role in the 
studies of multiple M2-branes \cite{BLG}. One may wonder 
if \eq{xxx-x} for a quantum
Nambu bracket may also play some role in the physics of M-branes. 
It is also
interesting  to understand better the mathematical properties of
\eq{xxx-x}, e.g. its center and representations, 
and to construct quantum field theory of this
kind of quantum space as well.

Finally we emphasise that the quantum Nambu bracket we considered in this paper 
should not be confused with the quantization of the Nambu-Poisson
bracket. The Nambu-Poisson bracket was introduced by
Takhtajan \cite{tak} and obeys the fundamental identity. Its
quantization is  a very hard mathematical 
problem and so far there is no satisfactory solution to it. It is 
an
interesting question as to
whether a quantum Nambu-Poisson bracket, or
higher n-ary
structures which are constrained by a fundamental identity
\cite{n-ary}, appears in the description of 
quantum geometry in string theory. 
On the other hand, it is very possible that one could
generalize the consideration of this paper by considering
higher form flux and find a higher $N$-bracket (completely
antisymmetrized sum of $N$ elements)
quantum geometry in string theory.

\section*{Acknowledgements}
We are grateful to Satoshi Ito, Sanjaye Ramgoolam, Harold
Steinacker  and Pichet Vanichchapongjaroen 
for discussions and particularly to Pei-Ming Ho and Douglas Smith
also for reading the draft of the paper and giving us many 
useful comments.
CSC wishes to thank
the participants of Corfu 2011 Summer Workshop on Noncommutative Field
Theory and Gravity, where some of the results of the paper were
presented,   
for  comments.
The work of CSC and GS is partially supported by STFC.

\end{document}